# Nanopores in atomically thin 2D nanosheets limit aqueous ssDNA transport

Alex Smolyanitsky*

*Applied Chemicals and Materials Division, National Institute of Standards and Technology, Boulder, Colorado 80305, USA*

Binquan Luan†

*IBM Thomas J. Watson Research Center, Yorktown Heights, New York 10598, USA*
(Dated: August 12, 2021)

Nanopores in 2D materials are highly desirable for DNA sequencing, yet achieving single-stranded DNA (ssDNA) transport through them is challenging. Using density functional theory calculations and molecular dynamics simulations we show that ssDNA transport through a pore in monolayer hexagonal boron nitride (hBN) is marked by a basic nanomechanical conflict. It arises from the notably inhomogeneous flexural rigidity of ssDNA and causes high friction *via* transient DNA desorption costs exacerbated by solvation effects. For a similarly sized pore in bilayer hBN, its self-passivated atomically smooth edge enables continuous ssDNA transport. Our findings shed light on the fundamental physics of biopolymer transport through pores in 2D materials.

Nanopore-based sensing of DNA nucleotides is becoming essential to human genome sequencing [1] and the recently proposed DNA storage technologies [2]. After decades of extensive studies, protein-nanopore-based DNA sequencing is poised to become a low-cost, high-throughput complement or possibly replacement for the existing technologies, including the now-ubiquitous dye sequencing [3] and the Sanger method [4]. In the meantime, the use of solid-state nanopores for sequencing remains challenging. Despite significant merits, such as chemical and mechanical robustness, solid-state nanopores suffer from poorly controllable surface properties, ultimately reducing their usability [5, 6]. In order to achieve solid-state nanopores with functionalities similar to those in transmembrane proteins (*e.g.*, holding a single nucleotide inside the pore at any given time during transport), various two-dimensional (2D) materials have been proposed as host membranes [7–10]. Specifically, electrophoretic transport of DNA through graphene- and hexagonal boron nitride (hBN)-based nanopores has been studied extensively. To date, most cases of successful transport have been demonstrated with flexurally stiff double-stranded DNA (dsDNA) [7, 9, 10]. Continuous transport of significantly more flexible single-stranded DNA (ssDNA) required for sequencing, however, has proven to be far more difficult [11], partly due to fundamental limitations arising from non-specific interactions between the DNA bases in and around nanopores [5, 8, 12]. Recent advances in fabrication have yielded atomically sculpted nanopores in molybdenum disulfide (MoS$_2$) monolayers [13], graphene [14], and monolayer hBN [14, 15]; yet, all such pores fundamentally feature an atomically thin region that must be traversed by ssDNA during translocation.

To inform our basic understanding of biopolymer transport through 2D nanopores, we demonstrate that such pores are incompatible with the basic nanomechanics of ssDNA and this incompatibility must be accounted for when designing nanoscale apertures with predictable transport properties. To demonstrate a case unencumbered by this incompatibility, we present low-friction ssDNA transport through a longer nanopore with an atomically defined toroidal surface. Our focus is on the fundamental mechanisms responsible for the transport barriers rather than the low-barrier pores. In fact, depending on the sequencing mechanism, slow ssDNA translocation may be desirable [16, 17], which can be controllably achieved under a smaller bias or with a solvent viscosity gradient [13].

We investigate ssDNA transport through similarly sized hexagonal nanopores in monolayer hBN and in an AA'-stacked hBN bilayer. Using density functional theory (DFT) calculations and density functional theory molecular dynamics (DFTMD), we show that the optimized structure of the bilayer-based pore features covalent B-N fusing between layers, resulting in an atomically smooth toroidal edge that is stable at room temperature. Using classical all-atom molecular dynamics informed by our DFT calculations, we demonstrate that in the monolayer case the hBN-adsorbed ssDNA traverses the pore edge akin to a bicycle chain simultaneously bending and sliding around a solid obstacle that is smaller than the interlink spacing. Consequently, continuous sliding is shown to be impossible and thus the ssDNA is caused to transiently desorb at the pore edge, perturbing the hydration layer near the pore surface. Overall, this process results in considerable transport-opposing energy barriers. In contrast, the bilayer-based nanopore features an overall more transport-conducive edge, which enables nonequilibrium low-friction sliding of ssDNA.

The ssDNA translocation results presented here were obtained using all-atom molecular dynamics (MD) simulations. The simulation cell was a 6-nm-tall hexagonal prism (cell vectors defined by $a = b = 5$ nm, $c = 6$ nm, $\alpha = \beta = 90°$, $\gamma = 60°$), containing the hBN membrane, a single strand of poly(dT$_{20}$), and an explicit aqueous 0.5M KCl solution ($\sim$16000 particles total), as shown schematically in Figs. 1a (monolayer hBN) and 4a (bilayer hBN). Periodic boundaries were imposed in the $XYZ$-directions, which included a periodic ssDNA molecule similar to earlier work [18, 19]. The ssDNA molecule was sized appropriately so as to freely adsorb on both sides of the membrane, resulting in roughly 12 nucleotides adsorbed on its *cis* side. The DNA model was AMBER-based [20], while the water model was TIP4P [21, 22]. The non-bonded portion of the nanoporous hBN models was set up according to recent work [23]. The membrane atoms were harmonically restrained in the $XYZ$-directions (spring constant of 500 N/m) around their equilibrium positions, which were obtained from the DFT calculations (see below). All MD simulations were carried out using the GPU-accelerated Gromacs 2018.1 software [24, 25]. Before production simulations, all systems underwent at least 10 ns of relaxation in a semi-isotropic NPT relaxation (cell size constant in-plane) at $T = 300$ K and $P = 0.1$ MPa with a time-step of 1 fs. All production simulations were carried out in the NVT ensemble ($T = 300$ K) using a time-step of 2 fs. Unless stated otherwise, all DNA translocations across the pores were initiated by a uniform external electric field $E$, which we report in the form of effective transmembrane voltage $V = E \times h$ ($h = 6$ nm is the cell height). The partial atomic charges and equilibrium positions used in our MD simulations were obtained using DFT calculations set up according to earlier work [23]: Perdew, Burke and Ernzerhof (PBE) exchange functional [26], Gaussian plane-wave pseudopotentials [27], and DZVP basis sets for boron and nitrogen atoms [28]. The pre-DFT starting structure of the bilayer-based nanopore did not feature any covalent B-N bonding between the layers, similar to a system studied earlier [29]. The total electric charge of all DFT-optimized structures was set to zero. The atomic charges were calculated according to the Blöchl scheme [30]. The ground-state DFTMD simulations to test room-temperature stability of the fused bilayer-based pore were carried out for 1 ps in vacuum (NVT ensemble, $T = 300$ K maintained using a velocity-rescaling thermostat) with a time step of 1 fs. All DFT computations were performed using the CP2K software package [31].

The results for ssDNA translocation through a hexagonal pore in monolayer hBN are provided in Fig. 1b, where cumulative fluxes of nucleotides are shown as a function of simulated time for various values of $V$. Importantly, we observe the presence of a threshold bias, below which ssDNA transport ceases, suggesting a significant transport-opposing barrier. It is overcome at $V = 2.4$ V when the vertical electrostatic forces on the charged phosphate groups of ssDNA are sufficiently large to defeat static friction in the system. At such a large bias, transport onset is likely assisted by field-induced desorption due to reorientation of bases (dipole moment $p \sim 4.4$ Debye) from being parallel to the hBN surface. Estimated in earlier work [32], the binding free energy of a nucleobase is $\Delta G_b \sim 33$ kJ/mol, for a degree of desorption requiring $V \geq 1.6$ V so that $pV/L \geq \Delta G_b$ ($L \sim 0.4$ nm is the effective thickness of the hBN monolayer), consistent with $V = 2.4$ V observed here. Finally, during transport the ssDNA underwent a series of ratchet-like motions (see the clearly resolved equally spaced steps in Fig. 1b), as discussed later.

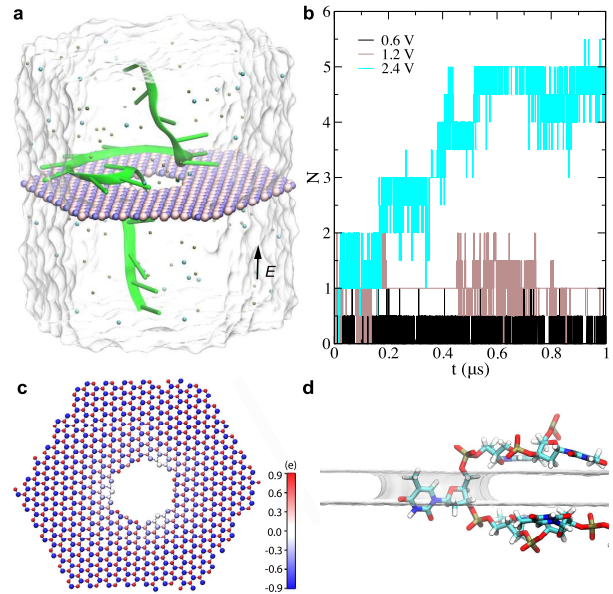

FIG. 1: ssDNA transport through a nanopore in an hBN monolayer. a) MD-simulated system. The ssDNA (colored in green) is in the cartoon representation; boron and nitrogen atoms in the membrane are shown as pink and blue spheres respectively; in the solution, $K^+$ and $Cl^-$ ions are in tan and cyan respectively, and water is shown as a transparent surface. b) Cumulative flux of ssDNA *vs* time at various bias voltages. c) Charge distribution of all membrane atoms. d) A side-view snapshot of ssDNA traversing the monolayer-based nanopore at $V = 0.6$ V. The hBN membrane is in the molecular surface representation, while the ssDNA molecule is in the stick representation.

From the simulation trajectories, we observe the representative transiently trapped state of translocating ssDNA as shown in Fig. 1d. Namely, a nucleotide resides inside the nanopore, desorbed from its surface and surrounded by water, while the two neighboring nucleotides are adsorbed on the *trans* and *cis* sides of the hBN monolayer, respectively. This suggests that upon transiting a monolayer-based pore a nucleotide must first desorb from the *cis* hBN surface, enter the water environment inside the pore (Fig. 1d), and finally readsorb on the *trans* side. This transition is expected to impose barriers associated with desorption and redistributing water surrounding the pore edge. This process, although seemingly inconsistent with the usual assumption of ssDNA's near-negligible flexural ridigity (given its persistence length of $\sim 2$ nm [33]), is easy to understand. In the situation above, it must be taken into account that the flexural rigidity of a single nucleotide is considerably higher than that between the nucleotides, separated by a finite distance $l_d \sim 0.68$ nm [33]. This results in a locally stiffened freely-jointed structure reminiscent of a bicycle chain, which plays a role when traversing a curved region as sharp as the edge of an atomically-thin monolayer. This piece-wise flexural inhomogeneity can be demonstrated in various ways, including directly inspecting the ssDNA MD model [20], known to yield flexural data in agreement with experiments [34] or by considering a typical ssDNA coarse-grain (CG) model [35]. Here we briefly illustrate it by considering selected angular fluctuations of a periodic poly(dT$_{12}$) strand in 0.5 M of aqueous KCl. Consider the CG sketch in the inset of Fig. 2: the linkages close to half-length of each grain (a nucleotide in this case) at the C$_4$ carbons are denoted joint 1, while the phosphate hinges between the grains separated by $l_d$ are denoted joint 2 at the P atoms. The corresponding angles of interest are formed by the C$_4$-P-C$_4$ and P-C$_4$-P atomic triplets. To exclude fast angular motions extraneous to the CG view, each set of atomic positions used to calculate the angles was obtained from an average of 20 raw position frames, corresponding to angles calculated and recorded every 0.4 ns. Eight individual angles for each triplet were considered, thus yielding a total of $10^4$ samples to build each histogram. From the fluctuation-dissipation theorem, the ratio between the corresponding flexural stiffness values is then calculated from the corresponding angular variances as $K_1/K_2 = \sigma_2^2/\sigma_1^2 = \sigma_{C_4PC_4}^2/\sigma_{PC_4P}^2 \approx 3.62$. Regardless of some multistability of angles (caused by stacking interactions) shown in Fig. 2, the local flexural heterogeneity is evident. Consequently, when vertically traversing an edge with curvature $> 1/l_d$, desorption at the edge is energetically preferred to extreme bending, causing behaviors shown in Figs. 1b,d.

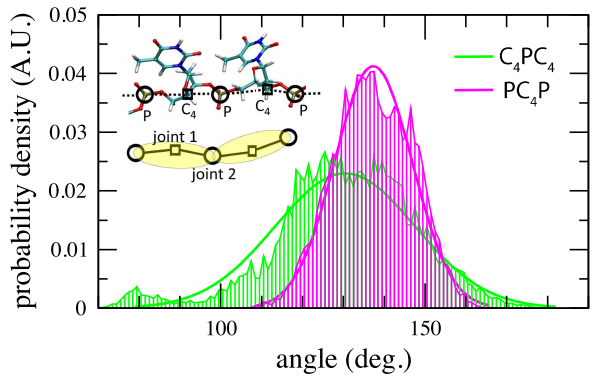

FIG. 2: Angle distributions (vertical bars and continuous curves show the raw histograms and the corresponding Gaussian fits, respectively), obtained from a $0.5\text{-}\mu s$-long MD simulation of a periodic poly(dT$_{12}$) in aqueous KCl. The corresponding sketch of ssDNA coarse-graining is provided in the inset.

To investigate a case without the interfacial discontinuity described above, we consider a bilayer-hBN-based nanopore, as shown in Figure 3a. Upon DFT optimization of the bilayer-based system (1492 atoms total), covalent B-N bonding occurs between the layers, as shown in Figs. 3b,c, resulting in a smooth toroidal-like inner pore surface, similar to "lip-lip" fusing reported earlier [36]. In the self-passivated region, the partial atomic charges near the pore were $Q_B = +0.62$e and $Q_N = -0.62$e, reduced from the bulk hBN values of $Q_B^{\text{bulk}} = +0.9$e and $Q_N^{\text{bulk}} = -0.9$e [23] (see charge distributions in Fig. 3d,e). The electron density distribution at the edge naturally inherited the polarity-alternating pattern of bulk hBN, unlike that for the monolayer-based pore, which possesses water-exposed radially dipolar edges (see Fig. 1c). As shown later, the hydration layer stability around the bilayer-based pore is reduced considerably in comparison with the monolayer-based case. The stability of the fused pore was confirmed in a room-temperature DFTMD simulation: Figure 3f shows that the potential energy of the system increased by $\Delta E \sim 37.6$ kJ/mol from that corresponding to the starting statically optimized structure and remained nearly constant throughout the DFTMD simulation.

After feeding the atomic positions and charges from the DFT calculations into the classical MD setup, we simulated aqueous ssDNA translocation through the bilayer-based nanopore. Figure 4a shows the simulated system. Upon the initial equilibration without a biasing field, the majority of nucleotides resided on the *cis* surface of the hBN bilayer. At a biasing voltage of 0.6 V, the ssDNA molecule translocated in about 0.3 $\mu$s. Note that due to the periodic set-up of the ssDNA molecule, only 12 out of 20 nucleotides initially resided on the *cis* surface and thus could transit the pore, as shown in Fig. 4b. At a lower bias of 0.3 V, the ssDNA molecule remained able to transit the pore, but the overall transport speed was reduced 3-4 times. At an even lower bias of 0.15 V, we observed translocation of only three nucleotides within the simulated time, suggesting a critical bias of order 0.15 V, an order of magnitude lower than that for the monolayer-based pore (2.4 V). As shown in Fig. 4c, during transport, ssDNA nucleotides remained adsorbed on the smooth pore edge due to its compatibility with the internucleotide spacing. However, as will be shown later, even in the event of transient desorption, given the bulk-like edge electrostatics, the surrounding water layer is considerably less stable than that around the monolayer-based pore or possibly any monolayer-based pore featuring water-exposed locally charged edges.

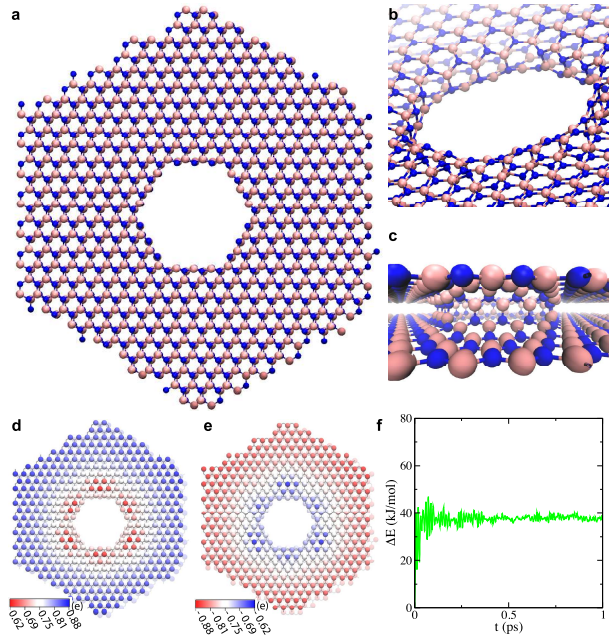

FIG. 3: DFT-based calculations for the nanopore in an hBN bilayer. a) Top view of the porous membrane. b) Tilted and enlarged view of the pore area (after static optimization). c) Side and perspective views of the post-optimization pore region. d) Charge distribution of all B atoms. e) Charge distribution of all N atoms. f) Change in potential energy of the system as a function time as obtained in a DFTMD simulation ($T = 300$ K), relative to the starting atomic configuration.

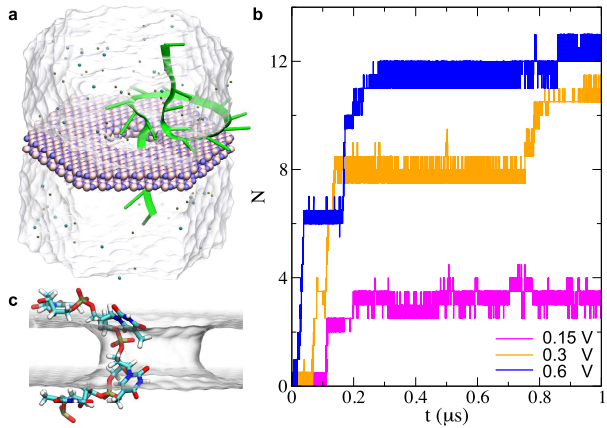

FIG. 4: ssDNA transport through a nanopore in an hBN bilayer. a) MD simulation system. The illustration scheme is identical to that in Figure 1. b) Cumulative flux of ssDNA *vs* time at various bias voltages. c) A snapshot of ssDNA near the pore at $V = 0.6$ V. The bilayer hBN membrane is in the molecular surface representation, while the ssDNA molecule is in the stick representation.

For a more quantitative analysis, we performed *unbiased* MD simulations of both pores, starting with ss-DNA conformations shown in Figs. 1d and 4c. For the bilayer-based case, the in-pore region of ssDNA (including the nucleotides adsorbed above and below) remained continuously adsorbed. From a 40-ns-long MD simulation, we obtained the average potential energies between each individual nucleotide and the hBN bilayer, shown in Fig. 5a. The two nucleotides roughly inside the pore are labeled $B_0$ and $B_1$; the two nucleotides on the *cis* side of the bilayer are labeled as $B_{-2}$ and $B_{-1}$; the two nucleotides on the *trans* side are labeled as $B_2$ and $B_3$. The results in Fig. 5a demonstrate that for all nucleotides in the vicinity of the bilayer-based pores had comparable interaction energies for all nucleotides, confirming adsorption throughout the pore region. However, as shown in Fig. 5a, the average interaction energy for the nucleotide inside the monolayer-based pore (Fig. 1d) is significantly smaller (less negative) than those for the neighboring nucleotides above and below, confirming desorption. For further illustration, the inset in Fig. 5a provides a side-view sketch of the sliding characters in the monolayer and bilayer case. The blue shading around each pore edge corresponds to the strength of the local pore-water interactions. As posited earlier, the pore-water interactions should be stronger for the monolayer-based pore, which further contributes to the corresponding transport barrier. We tested this hypothesis in two different ways. First, we calculated the corresponding time-averaged energies from the unbiased MD simulations used to obtain the data in Fig. 5a above. The average pore-water energy was $-605.7\pm8.6$ kJ/mol and $-488.2\pm6.9$ kJ/mol for the monolayer and bilayer case, respectively. The energy difference of almost 120 kJ/mol supports our hypothesis. Next, to observe the effect of dynamic water displacement by translocating nucleotides, we performed 200-ns-long steered MD simulations, in which the ssDNA molecule was pulled by harmonic springs through the monolayer and bilayer-based pore at a constant rate of 0.2 Å/ns, resulting in translocation of six bases. Throughout the simulated time, we recorded the pore-water interaction energy $U(t)$ as a function of simulated time, shown in the inset of Fig. 5b. If each translocation causes a perturbation in $U(t)$, one expects a peak at ~30 MHz in the corresponding frequency spectrum. As shown in Fig. 5b, a prominent peak is indeed observed in the monolayer case, strongly evidencing that transient nucleotide desorption is accompanied by a considerable perturbation of the surrounding water shell.

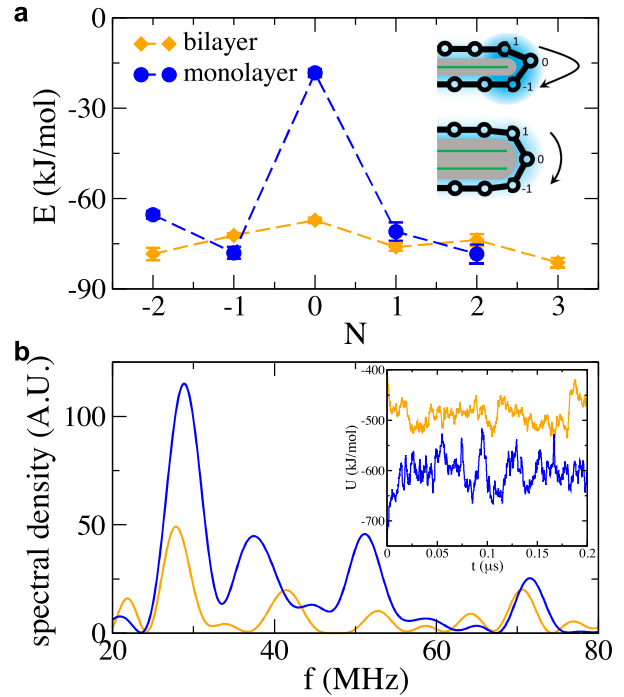

FIG. 5: Nucleotide-pore and water-pore interaction energies. a) Potential energy of interaction between individual ssDNA nucleotides and the pore vicinity for monolayer hBN (blue) and bilayer hBN (orange). The inset shows a sketch of edge traversal for monolayer- or bilayer-based pore. b) Frequency spectra of $U(t)$, as obtained in constant-rate pulling simulations of ssDNA; the $U(t)$ data is provided in the inset.

In summary, we have demonstrated that aqueous ssDNA transport through nanopores in atomically thin monolayers is fundamentally limited by a local nanomechanical discontinuity and the resulting desorption energy barrier, which arise when the effective pore edge radius is smaller than the internucleotide spacing. As a result, continuous translocation of ssDNA is precluded by significant transport barriers caused by transient desorption of nucleotides near the pore edge. This process is shown to be exacerbated by the perturbation of the local water shell. In contrast, a pore formed in AA'-stacked hBN bilayer and featuring a toroidal bulk-like edge surface is shown to be nanomechanically compatible with ssDNA, yielding continuous low-friction transport. Our observations are unlikely to be ssDNA-specific and are in principle applicable to any biopolymers that adsorb on 2D nanosheets prior to transport in aqueous environment (*e.g.*, denatured proteins). In addition to demonstrating a fundamental limitation of biopolymer transport through nanopores in 2D nanosheets along with an example solution, the results presented here should inform nanoscale aperture design considerations toward making functional solid-state nanopores for DNA and protein sequencers.

## ACKNOWLEDGMENTS

B.L. gratefully acknowledges support from the HCLS division. A.S. gratefully acknowledges support from the Materials Genome Initiative.

## REFERENCES

* Corresponding author: alex.smolyanitsky@nist.gov

† Corresponding author: bluan@us.ibm.com